\documentstyle[psfig]{nature_local}

\title{Nuclear-powered millisecond pulsars and the\\
       maximum spin frequency of neutron stars}

\author{Deepto~Chakrabarty\affiliation{Department of Physics and Center 
           for Space Research, Massachusetts Institute of Technology, 
           Cambridge, Massachusetts 02139, USA}\affiliation{Kavli
           Institute for Theoretical Physics, University of California,
           Santa Barbara, California 93106, USA},
        Edward~H.~Morgan$^*$,
        Michael~P.~Muno$^*$,\hfill\break
        Duncan~K.~Galloway$^*$,
        Rudy~Wijnands\affiliation{School of Physics and Astronomy, University 
           of St. Andrews, North Haugh, St. Andrews, Fife KY16~9SS, UK},
        Michiel~van~der~Klis\affiliation{Astronomical Institute ``Anton
            Pannekoek'' and Center for High-Energy Astrophysics, University 
            of Amsterdam, Kruislaan~403, 1098~SJ Amsterdam, The Netherlands},
        \hfill\break
        \& Craig~B.~Markwardt\affiliation{Department of Astronomy, University 
            of Maryland, College Park, Maryland 20742,
           USA}\affiliation{Laboratory for High Energy Astrophysics,
           NASA Goddard Space Flight Center, Greenbelt, Maryland 20771, USA}
}

\headertitle{}
\mainauthor{}

\summary{\noindent
Millisecond pulsars are neutron stars that are thought to have been spun-up
by mass accretion from a stellar companion\cite{acr+82}.  It is
unknown whether there is a natural brake for this process, or if it
continues until the centrifugal breakup limit is reached at
submillisecond periods.  Many neutron stars that are accreting mass
from a companion star exhibit thermonuclear X-ray bursts that last
tens of seconds, caused by unstable nuclear burning on their
surfaces\cite{sb03}.  Millisecond-period brightness oscillations
during bursts from ten neutron stars (as distinct from other rapid
X-ray variability that is also observed\cite{van00,wvh+03}) are 
thought to measure the stellar spin\cite{sb03,szs+96}, but direct
proof of a rotational origin has been lacking.  Here, we report the
detection of burst oscillations at the known spin frequency of an
accreting millisecond pulsar, and we show that these oscillations always
have the same rotational phase.  This firmly establishes burst
oscillations as nuclear-powered pulsations tracing the spin of
accreting neutron stars, corroborating earlier
evidence\cite{szs+96,sm02}.  The distribution of spin frequencies
of the 11 nuclear-powered pulsars cuts off well below the breakup
frequency for most neutron star models, supporting theoretical
predictions that gravitational radiation losses can limit accretion
torques in spinning up millisecond pulsars\cite{wag84,bil98,aks99}.}  
\dates{19 March 2003}{21 May 2003}

\begin{document}
\maketitle

The millisecond oscillations observed during X-ray bursts are not
perfectly coherent, but usually drift in frequency by several hertz
over the course of a burst, generally reaching an asymptotic maximium
frequency that is repeatable in a given neutron star\cite{sb03}.  This
frequency drift has been interepreted as arising from angular momentum
conservation in a decoupled surface burning layer that expands and
contracts during the burst, so that the asymptotic frequency is the
stellar spin frequency\cite{sjg+97,cb00}.  A puzzle in this picture is
why the oscillation persists late in the burst, well after the nuclear
burning has spread over the entire star.  Also, in most of these
neutron stars, unexplained pairs of kilohertz quasi-periodic
oscillations (kHz QPOs) are also observed in the non-burst X-ray
emission, with the QPO separation frequency approximately equal to
either the burst oscillation frequency or half this value\cite{van00},
posing a further puzzle (see ref.~\pcite{wvh+03} for new insight into
this problem).

We have observed the transient X-ray source
SAX~J1808.4$-$3658, which has been detected in four outbursts since
its discovery (September 1996, April 1998, January 2000, and October
2002), each lasting several weeks.  It was previously established
that the source is a weakly magnetized ($<10^{10}$~G), rapidly
rotating (401~Hz) accreting neutron star in a 2~hr low-mass X-ray
binary\cite{ihm+98,wv98,cm98}.  We observed it during its 2002 X-ray
outburst for about 700,000 seconds between 15 October and 26 November
using the Proportional Counter Array (PCA) on the Rossi X-Ray Timing
Explorer (RXTE).  As in the two previous outbursts observed by RXTE,
persistent 401~Hz accretion-powered X-ray pulsations\cite{wv98}
modulated by the 2~hr binary orbit\cite{cm98} were detected throughout
our observations, with a fractional r.m.s. amplitude of 3--5\%. The
pulsar spin frequency derived from these data was approximately
400.97521~Hz at the start of the outburst, with a mean spin-down rate
of about $2\times 10^{-13}$ Hz~s$^{-1}$; a detailed discussion of the
pulsar's spin evolution will be presented elsewhere.  Four
thermonuclear X-ray bursts were also detected on 15, 17, 18, and 19
October; these are among the brightest X-ray bursts ever observed by
RXTE from any neutron star.  Previous analysis of observations from the 1996
outburst of SAX J1808.4$-$3658 with the BeppoSAX satellite yielded a
marginal detection of a 400$\pm$2~Hz oscillation during a bright X-ray
burst\cite{ick+01}. 

\begin{figure}[t]
\centerline{\psfig{file=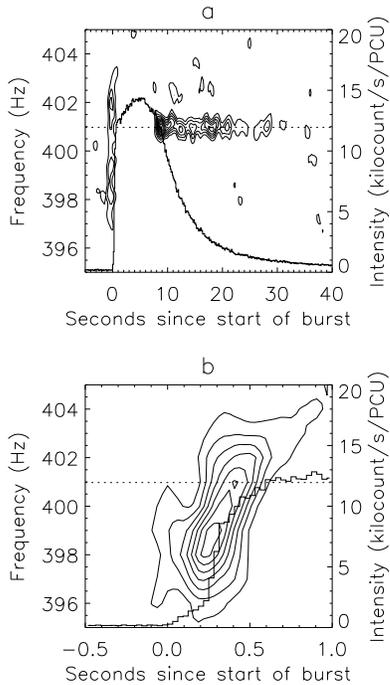,width=2.1in}}
\caption{Dynamic power spectra of millisecond oscillations in an X-ray
  burst on 18 October 2002 from SAX J1808.4$-$3658. (a) We analyzed the
  2--60~keV PCA data binned at 122~$\mu$s resolution with no energy
  resolution, with the bin times corrected to the binary
  center-of-mass frame.  We computed overlapping, oversampled Fourier
  power spectra of 2~s duration spaced at 0.25~s intervals.  The
  contours show Fourier power levels as a function of frequency and
  time, and the solid histogram shows the X-ray intensity of the burst.  The
  horizontal dotted line shows the (known) pulsar spin frequency.  A
  rapidly drifting oscillation is detected during the burst rise, and
  a stationary oscillation at nearly the spin frequency is
  detected in the burst tail.  The $n$-th contour level has a
  single-trial probability of $0.02^n$ of occurring due to noise
  fluctuations. (b) A close-up view of the burst rise, computed using
  power spectra of 0.25~s duration at 0.03125~s intervals.  The smooth
  frequency drift is more clearly apparent here.
  }
\end{figure}
Strong millisecond oscillations around 401~Hz were clearly detected in
all four X-ray bursts in the October 2002 data, with a fractional
r.m.s. amplitude of 3--5\% and very similar characteristics in each burst
(Fig. 1). First, a rapidly drifting oscillation was detected during
the burst rise in the 397--403~Hz range.  Second, no oscillations were
detected ($<$1.5\% amplitude) during the peak luminosity phase of the
burst, when the photosphere is driven away from the star by radiation
pressure; this behavior is typical of other burst oscillation sources
as well\cite{mcg+02}. Finally, a strong oscillation reappeared during 
the cooling phase of the burst at a constant frequency nearly equal to
the pulsar spin frequency, but exceeding it by an average of $6\pm
1$~mHz in the three best-sampled bursts.  The oscillation amplitudes
are comparable to those of the persistent (non-burst)
accretion-powered pulsations even though the burst emission is
considerably brighter than the persistent emission, indicating that
the burst flux itself must be pulsed. The burst oscillations are
sinusoidal in shape, with a 2$\sigma$ upper limit of $<$0.5\% on the
fractional r.m.s. amplitude of an 802~Hz harmonic, compared to a
measured mean harmonic amplitude of 0.4\% for the persistent
pulsations. 

The observed frequency drift demonstrates that this is a similar
phenomenon as the burst oscillations observed in ten other neutron stars,
although the oscillations in this neutron star have some unusual traits.
The frequency drift is among the largest observed in any
neutron star\cite{wsf01,gcm+01} ($\Delta\nu\approx$5~Hz,
$\Delta\nu/\nu\approx$1\%).  Also, the drift time scale is an order of
magnitude faster than in the other neutron stars\cite{mcg+02}, and the
maximum oscillation frequency is reached during the burst rise,
inconsistent with angular momentum conservation in a cooling,
contracting shell.  In fact, the oscillation overshoots the spin
frequency during the burst rise.  The rapid frequency drift may be an
indication that SAX J1808.4$-$3658 has a stronger magnetic field than
the other neutron stars, since a sufficiently strong field will suppress
rotational shearing in the burning layer\cite{cb00} and may act as a
restoring force.

This magnetic field argument is particularly appealing given that SAX
J1808.4$-$3658 is the only burst oscillation source, and one of only 4
neutron stars in low-mass X-ray binaries out of over 70, that shows persistent
pulsations in its non-burst emission\cite{mss+02,gcm+02,mss03}.  The
absence of persistent pulsations in most of these neutron stars suggests
that they lack a sufficiently strong field for the accretion flow to be
magnetically channeled.  Indeed, it has been proposed that the absence
of persistent millisecond pulsations from most neutron stars in low-mass X-ray
binaries is due to diamagnetic screening of the neutron star magnetic field by
freshly accreted material for the typical range of mass accretion
rates\cite{czb01} ($\dot M\geq 10^{-10} M_\odot$~yr$^{-1}$, where
$M_\odot$ is the solar mass).  In this context, we note that all four
persistent millisecond X-ray pulsars lie at the low end ($\dot M\sim
10^{-11} M_\odot$~yr$^{-1}$) of the mean $\dot M$ distribution for
low-mass X-ray binaries.  If this explanation is correct, then most
burst oscillation sources are unlikely to show persistent pulsations.

\begin{figure}[t]
\centerline{\psfig{file=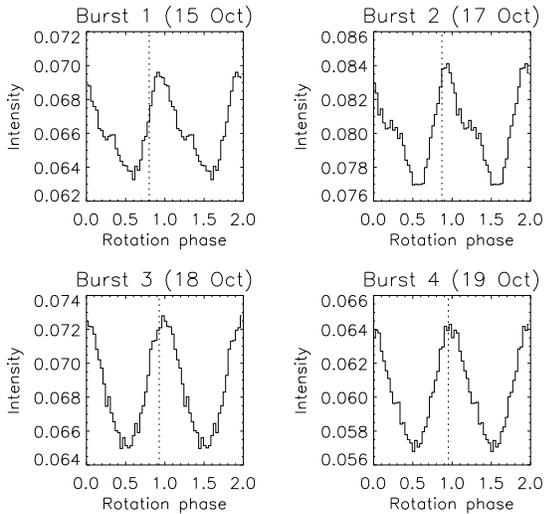,width=3in}}
\caption{Relative phase of the burst tail oscillations in SAX
  J1808.4$-$3658 and the persistent accretion-powered pulsations just
  prior to the burst, for all four bursts.  The histograms show the
  profile of the persistent pulsations, and the dashed lines indicate
  the rotational phase of the burst tail oscillations.  The
  oscillations all have nearly the same rotational phase and slightly
  lead the persistent pulsations.} 
\end{figure}
We can measure the rotational phase of the oscillations in the burst
tails.  For each burst, we determined a rotational ephemeris using the
persistent (accretion-powered) pulsations in the data for a few
thousand seconds prior to the burst and compared this to the phase
(epoch of maximum intensity) of the oscillation in the burst tail.
The burst tail oscillations all had the same rotational phase within
$\pm$6\% and were roughly phase-aligned with the non-burst pulsations,
leading them by an average of 11\% (Fig. 2).  These measurements 
will help resolve the puzzle of why oscillations persist in the burst tail,
when the nuclear burning has presumably enveloped the entire star. 
The phase locking indicates that the oscillations are associated with
the stellar surface and suggests that the emitting ``hot spot'' has a
nearly fixed orientation with respect to the pulsar's magnetic axis.
The 11\% offset is consistent with the phase drift accumulated over the
duration of the burst tails due to the slight ($<$0.002\%) frequency
difference of the burst oscillations and the non-burst pulsations, but
this requires an initial phase alignment in the burst tail followed by
slight motion of the hot spot\cite{slu02}.

Our observations clearly show that brightness oscillations in the
tails of thermonuclear X-ray bursts are a nearly exact tracer of
stellar spin, establishing these neutron stars as nuclear-powered
pulsars. (Other recent observations show that the kHz QPO frequency
separation in this neutron star is half the spin frequency\cite{wvh+03},
verifying that the burst oscillation traces the spin and not a harmonic.)  
This provides a powerful tool for studying the most rapidly
rotating neutron stars.  It is believed that accretion from a binary companion
is required in order to spin up old neutron stars to millisecond
periods\cite{acr+82}.  For sustained accretion in the absence of
another angular momentum sink, this process is limited only by the neutron star
breakup frequency, up to 3~kHz depending upon the nuclear equation of
state\cite{cst94,hlz99}.  It has been difficult to determine the
actual spin distribution of millisecond pulsars, since there have been
strong observational selection effects against detecting very short
period millisecond radio pulsars, and persistent millisecond X-ray
pulsars are rare.  However, RXTE has no significant selection effects
against detecting burst oscillations at frequencies well above 1~kHz,
so the nuclear-powered pulsars are ideal for probing the fast end of
the pulsar spin distribution.

The spin frequencies of the 11 nuclear-powered pulsars lie between
270~Hz and 619~Hz, and within that range are marginally consistent
with a uniform distribution.  However, the absence of
spins above 619~Hz is significant.  If we make the simple assumption
of a uniform spin distribution in the range [$\nu_{\rm low}$,
  $\nu_{\rm high}$] and consider a sample of $N$ pulsars with spin
frequencies $\nu_j$, then a Bayesian analysis yields the probability
density for $\nu_{\rm high}$ (within a normalization factor), 
\begin{equation}
  p(\nu_{\rm high}) = [\nu_{\rm high}-\min(\nu_j)]^{1-N} - \nu_{\rm
  high}^{1-N}
\end{equation}
for $\nu_{\rm high}\geq\max(\nu_j)$, where we have made an {\em a priori} 
allowance for a $0<\nu<3$~kHz range.  For the 11 observed spins, this
yields an upper limit of $\nu_{\rm high}<$760~Hz (95\% confidence)
on the maximum spin frequency, which is also consistent with the
fastest known (641~Hz) millisecond radio pulsar\cite{bkh+82}.  This is
well below the breakup frequency for nearly all models of rapidly
rotating neutron stars (except for those that have an extremely stiff equation
of state and contain a pion/kaon condensate in the core, and then only
for $M<1.5\,M_\odot$)\cite{cst94b}.  Some mechanism clearly acts to
halt pulsar spin-up while accretion is still active.   

Steady magnetic accretion torques should drive an accreting pulsar to
an equilibrium spin frequency that depends upon $\dot M$ and the
surface dipole magnetic field $B$.  Taking the $\dot M$ range observed
in the nuclear-powered pulsars and the disk-magnetosphere interaction
models relevant for weakly magnetic neutron stars\cite{pc99}, magnetic
spin equilibrium can account for the observed spin distribution if all
these systems have $B\sim 10^8$~G, consistent with the field inferred
in SAX J1808.4$-$3658 (ref. \pcite{pc99}) and in the millisecond radio
pulsars\cite{ctk94}.  However, fields this strong should be
dynamically important for the accretion flow, making it difficult to
understand the lack of persistent pulsations in the non-burst emission
of all the nuclear-powered pulsars other than SAX J1808.4$-$3658 and
instead suggesting a broader range of field strengths.  Alternatively,
several authors have shown that gravitational radiation can carry away
substantial angular momentum from accreting neutron stars, driven
either by the excitation of an $r$-mode instability in the neutron
star core\cite{wag84,aks99} or by a rotating, accretion-induced
crustal quadrupole moment\cite{bil98}. These losses can balance
accretion torques for the relevant ranges of $\nu$ and $\dot M$, and
the predicted gravitational radiation strengths are near the detection
threshold for planned gravitational wave interferometers, especially
in the case where an X-ray timing ephemeris is available\cite{bil03}.


\begin{thebibliography}{10}

\bibitem[Alpar {\it et~al.}<1>]{acr+82}
Alpar, M.~A., Cheng, A.~F., Ruderman, M.~A.  \& Shaham, J. A new class of radio
  pulsars, {\it Nature} {\bf 300}, 728--730 (1982).

\bibitem[Strohmayer \& Bildsten<2>]{sb03}
Strohmayer, T. \& Bildsten, L. in {\it Compact Stellar X-Ray Sources} (eds
  Lewin, W.~H.~G. \& van~der Klis, M.)  in press (astro--ph/0301544) (Cambridge
  U. Press, Cambridge, 2003).

\bibitem[van~der Klis<3>]{van00}
van~der Klis, M. Millisecond oscillations in X-ray binaries, {\it Ann. Rev.
  Astr. Astrophys.} {\bf 38}, 717--760 (2000).

\bibitem[Wijnands {\it et~al.}<4>]{wvh+03}
Wijnands, R. {\it et al.} Kilohertz quasi-periodic X-ray brightness
  fluctuations from an accreting millisecond pulsar, {\it Nature} {\bf }, in
  press (2003).

\bibitem[Strohmayer {\it et~al.}<5>]{szs+96}
Strohmayer, T.~E. {\it et al.} Millisecond variability from an accreting
  neutron star system, {\it Astrophys. J.} {\bf 469}, L9--L12 (1996).

\bibitem[Strohmayer \& Markwardt<6>]{sm02}
Strohmayer, T.~E. \& Markwardt, C.~B. Evidence for a millisecond pulsar in 4U
  1636$-$53 during a superburst, {\it Astrophys. J.} {\bf 577}, 337--345
  (2002).

\bibitem[Wagoner<7>]{wag84}
Wagoner, R.~V. Gravitational radiation from accreting neutron stars, {\it
  Astrophys. J.} {\bf 278}, 345--348 (1984).

\bibitem[Bildsten<8>]{bil98}
Bildsten, L. Gravitational radiation and rotation of accreting neutron stars,
  {\it Astrophys. J.} {\bf 501}, L89--L93 (1998).

\bibitem[Andersson, Kokkotas \& Stergioulas<9>]{aks99}
Andersson, N., Kokkotas, K.~D.  \& Stergioulas, N. On the relevance of the
  r-mode instability for accreting neutron stars and white dwarfs, {\it
  Astrophys. J.} {\bf 516}, 307--314 (1999).

\bibitem[Strohmayer {\it et~al.}<10>]{sjg+97}
Strohmayer, T.~E., Jahoda, K., Giles, A.~B.  \& Lee, U. Millisecond pulsations
  from a low-mass X-ray binary system in the Galactic center region, {\it
  Astrophys. J.} {\bf 486}, 355--362 (1997).

\bibitem[Cumming \& Bildsten<11>]{cb00}
Cumming, A. \& Bildsten, L. Rotational evolution during type I X-ray bursts,
  {\it Astrophys. J.} {\bf 544}, 453--474 (2000).

\bibitem[in~'t Zand {\it et~al.}<12>]{ihm+98}
in~'t Zand, J.~J. {\it et al.} Discovery of the X-ray transient
  SAX~J1808.4--3658, a likely low-mass X-ray binary, {\it Astr. Astrophys.}
  {\bf 331}, L25--L28 (1998).

\bibitem[Wijnands \& van~der Klis<13>]{wv98}
Wijnands, R. \& van~der Klis, M. A millisecond pulsar in an X-ray binary
  system, {\it Nature} {\bf 394}, 344--346 (1998).

\bibitem[Chakrabarty \& Morgan<14>]{cm98}
Chakrabarty, D. \& Morgan, E.~H. The two-hour orbit of a binary millisecond
  X-ray pulsar, {\it Nature} {\bf 394}, 346--348 (1998).

\bibitem[in~'t Zand {\it et~al.}<15>]{ick+01}
in~'t Zand, J.~J. {\it et al.} The first outburst of SAX J1808.4--3658
  revisited, {\it Astr. Astrophys.} {\bf 372}, 916--921 (2001).

\bibitem[Muno {\it et~al.}<16>]{mcg+02}
Muno, M.~P., Chakrabarty, D., Galloway, D.~K.  \& Psaltis, D. The frequency
  stability of millisecond oscillations in thermonuclear X-ray bursts, {\it
  Astrophys. J.} {\bf 580}, 1048--1059 (2002).

\bibitem[Wijnands, Strohmayer \& Franco<17>]{wsf01}
Wijnands, R., Strohmayer, T.  \& Franco, L.~M. Discovery of nearly coherent
  oscillations with a frequency of 567~Hz during type I X-ray bursts of the
  X-ray transient and eclipsing binary X1658$-$298, {\it Astrophys. J.} {\bf
  549}, L71--L75 (2001).

\bibitem[Galloway {\it et~al.}<18>]{gcm+01}
Galloway, D.~K., Chakrabarty, D., Muno, M.~P.  \& Savov, P. Discovery of a 270
  hertz X-ray burst oscillation in the X-ray dipper 4U 1916$-$053, {\it
  Astrophys. J.} {\bf 549}, L85--L88 (2001).

\bibitem[Markwardt {\it et~al.}<19>]{mss+02}
Markwardt, C.~B., Swank, J.~H., Strohmayer, T.~E., in~'t Zand, J.~J.~M.  \&
  Marshall, F.~E. Discovery of a second millisecond accreting pulsar: XTE
  J1751$-$305, {\it Astrophys. J.} {\bf 575}, L21--L24 (2002).

\bibitem[Galloway {\it et~al.}<20>]{gcm+02}
Galloway, D.~K., Chakrabarty, D., Morgan, E.~H.  \& Remillard, R.~A. Discovery
  of a high-latitude accreting millisecond pulsar in an ultracompact binary,
  {\it Astrophys. J.} {\bf 576}, L137--L140 (2002).

\bibitem[Markwardt, Smith \& Swank<21>]{mss03}
Markwardt, C.~B., Smith, E.  \& Swank, J.~H. XTE J1807$-$294, {\it IAU Circ.}
  {\bf }, No. 8080 (2003).

\bibitem[Cumming, Zweibel \& Bildsten<22>]{czb01}
Cumming, A., Zweibel, E.  \& Bildsten, L. Magnetic screening in accreting
  neutron stars, {\it Astrophys. J.} {\bf 557}, 958--966 (2001).

\bibitem[Spitkovsky, Levin \& Ushomirsky<23>]{slu02}
Spitkovsky, A., Levin, Y.  \& Ushomirsky, G. Propagation of thermonuclear
  flames on rapidly rotating neutron stars: extreme weather during type I X-ray
  bursts, {\it Astrophys. J.} {\bf 566}, 1018--1038 (2002).

\bibitem[Cook, Shapiro \& Teukolsky<24>]{cst94}
Cook, G.~B., Shapiro, S.~L.  \& Teukolsky, S.~A. Recycling pulsars to
  millisecond periods in general relativity, {\it Astrophys. J.} {\bf 421},
  L117--L120 (1994).

\bibitem[Haensel, Lasota \& Zdunik<25>]{hlz99}
Haensel, P., Lasota, J.~P.  \& Zdunik, J.~L. On the minimum period of uniformly
  rotating neutron stars, {\it Astr. Astrophys.} {\bf 344}, 151--153 (1999).

\bibitem[Backer {\it et~al.}<26>]{bkh+82}
Backer, D.~C., Kulkarni, S.~R., Heiles, C.~E., Davis, M.~M.  \& Goss, W.~M. A
  millisecond pulsar, {\it Nature} {\bf 300}, 615--618 (1982).

\bibitem[Cook, Shapiro \& Teukolsky<27>]{cst94b}
Cook, G.~B., Shapiro, S.~L.  \& Teukolsky, S.~A. Rapidly rotating neutron stars
  in general relativity: realistic equations of state, {\it Astrophys. J.} {\bf
  424}, 823--845 (1994).

\bibitem[Psaltis \& Chakrabarty<28>]{pc99}
Psaltis, D. \& Chakrabarty, D. The disk-magnetosphere interaction in the
  accretion-powered millisecond pulsar SAX J1808.4$-$3658, {\it Astrophys. J.}
  {\bf 521}, 332--340 (1999).

\bibitem[Camilo, Thorsett \& Kulkarni<29>]{ctk94}
Camilo, F., Thorsett, S.~E.  \& Kulkarni, S.~R. The magnetic fields, ages, and
  original spin periods of millisecond pulsars, {\it Astrophys. J.} {\bf 421},
  L15--L18 (1994).

\bibitem[Bildsten<30>]{bil03}
Bildsten, L. in {\it Radio Pulsars} (eds Bailes, M., Nice, D.~J.  \& Thorsett,
  S.~E.)  in press (astro--ph/0212004) (Astron. Soc. Pacific, San Francisco,
  2003).

\end{thebibliography}

\smallskip
\noindent {\small {\bf Acknowledgements.} We thank L. Bildsten,
F. Lamb, D. Psaltis, and S. Thorsett for useful discussions.  This
work was supported in part by NASA and the Alfred P. Sloan Foundation.
}

\medskip
\noindent {\small {\bf Competing interests statement.} The authors
  declare that they have no competing financial interests.}

\medskip
\noindent {\small {\bf Correspondence} and requests for materials should be
  addressed to D.C. (e-mail: deepto@space.mit.edu).} 

\end{document}